\documentclass[useAMS,usenatbib]{mn2e}
\usepackage{amsmath,amssymb}
\def\rmmat#1{{\hbox{\rm #1}}}
\def\rmscr#1{\rmmat{\scriptsize #1}}

\usepackage{graphicx} 
\usepackage{dcolumn} 
\usepackage{bm} 
\usepackage{psfrag}

\begin{document}

\title[Magnetospheric Birefringence]{Magnetospheric Birefringence Induces  Polarization Signatures in Neutron-Star Spectra}

\author[R. M. Shannon and J. S. Heyl]{R. M. Shannon\thanks{Current Address:  Department of Astronomy, 
Cornell University, Ithaca, New York, United States, 14853; email: ryans@astro.cornell.edu}
and J. S. Heyl\thanks{email: heyl@phas.ubc.ca; Canada Research Chair}\\
Department of Physics and Astronomy, University of British Columbia, Vancouver, British Columbia, Canada, V6T 1Z1}

\date{Accepted 2006 February 16.  Received 2006 February 15; in original form 2006 January 19 }

\pagerange{\pageref{firstpage}--\pageref{lastpage}} \pubyear{2006}

\maketitle
\label{firstpage}

\begin{abstract}
We study the propagation of polarization light through the
magnetosphere of neutron stars.  At intermediate frequencies (the
optical through the infrared), both the birefringence induced by the
plasma and by quantumelectrodynamics influence the observation
polarization of radiation from the surface of the neutron star.
Because these two processes compete in this regime, we find that
polarization observations can constrain the properties of the
neutron-star magnetosphere, specifically the total charge density.  We
calculate both the phase-resolved and the phase-averaged polarization
signatures induced by magnetospheric birefringence.
\end{abstract}


\section{Introduction}

The propagation of polarized radio waves away from neutron stars (NSs)
has been well understood for many years \citep{Barn86}.  At this
energy scale, it is reasonable to neglect the coupling of the light to
vacuum-based QED effects.  Lai and Ho
(2003ab)\nocite{Lai:2003,Lai:2003b} have discussed the mechanisms by
which polarized X-rays are affected by the atmospheres of NSs. X-ray
photons couple the greatest in the atmosphere as the plasma densities
and field strengths are sufficiently high.  Lai and Ho included
vacuum-based effects in their analysis.  Similar effects exist at
IR/optical wavelengths when propagation of the waves through the
magnetosphere (possessing a lower-density plasma and lower strength
magnetic field) is considered.  This paper presents a model and
results that show that there is a wavelength-dependent polarization
signature in the IR/optical wavelength as a result of interactions
between radiation and the magnetosphere.

For most radio pulsars emission from the magnetosphere itself rather
than the surface is thought to dominate the optical and near infrared
emission.  However, several radio pulsars exhibit optical radiation
from the surface ({\em e.g.} PSR~B0656+14, PSR~B0950+08 and
Geminga) \citep[\protect{\em e.g.}][]{2004A&A...417.1017Z}.  Radio pulsars are not the only
members of the neutron-star menagerie.  Tbe optical and near-infrared
emission from the radio quiet neutron stars such as RX~J1856.5-3754
and RX~J0720.4-3125\citep[\protect{\em e.g.}][]{2002ApJ...564..981P,2003A&A...408..323M} most
likely comes from the surface, and possibly the emission from
anomalous x-ray pulsars emerges from a surface layer heated by
magnetospheric currents \citep{2004astro.ph..8538T}.  With these
objects in mind let us proceed.

To make progress we will make several simplifying assumptions. 
First, we assume that the charge carriers both positive and negative
have the mass of an electron and that the plasma is cold.  A pair
plasma is necessarily hot. Second, we assume that the Goldreich-Julian
model is correct.  The Goldreich-Julian picture describes the
equilibrium charge density for a rotating neutron star with its
magnetic pole aligned (or anti-aligned) with the spin axis.  Here we
will assume that the two axes are not necessarily aligned (especially
in \S~\ref{sec:phase-integrations}).  Finally, the Goldreich-Julian
equilibrium even in its realm of applicability is unstable
\citep[see][]{2004AdSpR..33..542M}.  In spite of these deficiencies,
the Goldreich-Julian model provides a well-understood starting point
to demonstrate that observations of polarization of the surface
radiation from neutron stars may provide a window to diagnose the
properties of the plasma and to encourage further research to probe
more realistic models.

\section{Calculations : Polarization Evolution}

\citet{Kubo:1983} outline how the polarization of light is
affected by a birefringent and dichroic medium:
\begin{equation}
\frac{d \vec{s}}{dz} = \hat{\Omega} \times \vec{s}  + (\hat{T} \times \vec{s}) \times \vec{s} 
\label{propeqn}.
\end{equation}
This equation is valid for a medium with dispersion, emission and
absorption but not scattering.
The polarization is described in terms of normalized Stokes parameters
$\vec{s} = \left(s_1,s_2,s_3 \right) =$ (Q/I, U/I, V/I).  The vectors
$\hat{\Omega}$ and $\hat{T}$ are the 
birefringent and dichroic 
vectors.  We have extended the expressions of 
\citet{Heyl:2000} for a permeable medium to include the effects
of the plasma:
\begin{equation}
\hat{\Omega} = \frac{\omega}{2 c \sqrt{\epsilon_0}} \left (\begin{array}{c}
(\eta_E \epsilon_B - \epsilon_E \eta_B) \sin^2 \theta \cos  2 \phi \\
-(\eta_E \epsilon_B - \epsilon_E \eta_B) \sin^2 \theta \sin 2 \phi\\
g \cos \theta ( \epsilon_B + \frac{1}{2}(\eta_B - \epsilon_B)\sin^2 \theta)
\end{array} \right)
\label{birefringent},
\end{equation}
where $\omega$ is the angular frequency of the radiation;
$\theta$ is the angle between the direction of propagation and
the magnetic field; and $\phi$ is the angle between the projection of
the magnetic field onto the plane perpendicular to the direction of
propagation, and the component of the magnetic field in the direction
of the magnetic moment of the NS. $\eta_B$ and $\epsilon_B$
are components of the permeability tensor, and $\eta_E$ and
$\epsilon_E$ are components of the dielectric tensor, which are, to
first order in the fine-structure constant ($\alpha_f$):
\begin{eqnarray}
\eta_B &=& 1 + \frac{2}{15}\frac{\alpha_f}{\pi} \left(\frac{B}{B_Q}
\right)^2, \\
\epsilon_B &=& 1 + \frac{2}{45}\frac{\alpha_f}{\pi} \left(\frac{B}{B_Q} \right)^2,\\
\eta_E &=& 1 -v + \frac{1}{9}\frac{\alpha_f}{\pi} \left(\frac{B}{B_Q} \right)^2,\\
\epsilon_E &=& 1 - \frac{v}{1-u} - \frac{2}{45} \frac{\alpha_f}{\pi}\left(\frac{B}{B_Q} \right)^2,\\
g &=& \frac{v \sqrt{u}}{1 - u},\\
\epsilon_0 &=& \epsilon_E \epsilon_B \cos ^2 \theta + \frac{1}{2} \sin^2 \theta 
\left( \eta_E \epsilon_B + \epsilon_B \eta_B \right)
\end{eqnarray}
for $B \ll B_Q$ and $\hbar \omega \ll m_e c^2$ where $v =
\omega_p^2/\omega^2$, and $u = \omega_c^2/\omega^2$.
$\omega_c=eB/(m_e c)$ is the cyclotron frequency and 
\begin{equation}
\omega_p =
\sqrt {\frac{4\pi e^2n_e}{m_e}} =
\sqrt{2 \Omega \omega_c |\cos \theta'|}
\end{equation}
is the plasma frequency under the assumptions mentioned in the
introduction.

  To get the second equality we have assume the Goldreich-Julian
(1969) \nocite{Goldreich:1969} plasma density, $n_{GJ}=(\Omega
B\cos\theta')/(2\pi ec)$.  $\Omega=2\pi/P$ is the spin frequency of
the star and $\theta'$ is the angle between the rotation axis and the
local magnetic field.  $B$ is the strength of the magnetic field {\it
perpendicular to propagation direction} and $B_Q \approx 4.4 \times
10^{13}$~G is the critical QED field.  In the range of NSs modeled, it
was found that the dichroic vector had little effect on the
polarization, and was neglected for subsequent results presented here.

A fourth-order Runge-Kutta integrator with variable step size was used
to determine the evolution of the polarization vector from the surface
of the star to the light cylinder, along a Schwarzschild null
geodesic.  The polarization vector $\vec{s}$ was initially set to be
in the direction of the plasma-dominated birefringent vector, as the
initial polarization modes are determined by the plasma-rich
atmosphere of the NS.

We set the initial conditions by specifying the inclination angle
$\alpha$ (not to be confused with the fine structure constant
$\alpha_f$) between the optical axis and the magnetic axis, and the
angle $\beta$ between the plane containing the optical and magnetic
axes and the plane containing the photon trajectory.  For all the
calculations stellar radii $R = 10^6$ cm and mass $M = 1.4$
$M_{\odot}$ were considered.  In addition, we assumed that the NS's
magnetic field was dipolar.  The output of single integration is
similar to Figure~3 of \citet{Heyl:2003}.  The $s_1$ and $s_2$
polarizations change considerably as the photon travels away from the
NS.  At approximately 30 stellar radii, the polarization decouples.
The significance of the decoupling radius is discussed below.

\section{Results}

\subsection{Net Polarization}

\begin{figure}
\psfrag{s1}{$s_1$}
\psfrag{ 1e+13}{$10^{13}$}
\psfrag{ 1e+14}{$10^{14}$}
\psfrag{ 1e+15}{$10^{15}$}
\psfrag{30m}{$30\mu$m}
\psfrag{3m}{$3\mu$m}
\psfrag{0.3m}{$0.3\mu$m}
\includegraphics[width=7.6cm,height=4.6cm]{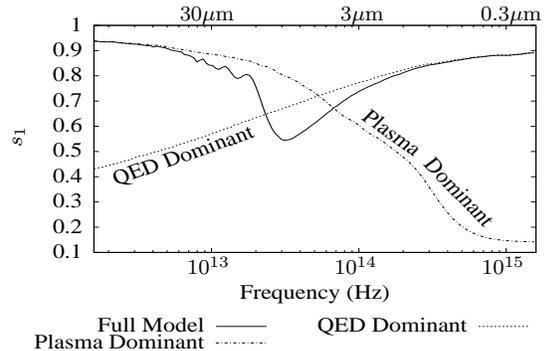}
\caption{\label{compare} Comparison of the three regimes for a neutron
star with a one-second spin period and a magnetic dipole moment of
$10^{31}$G~cm$^3$, pointing in a direction $60^\circ$ from the line of
sight.  The solid line represents the model containing both QED and
plasma-based effects, the dot-dashed line shows integrations
conducted using only plasma effects, and the short-dashed lines
represents integrations conducted with only vacuum-based effects.}
\end{figure}

The net polarization of the radiation emitted from the NS
was determined by integrating over many trajectories on the neutron
star image. We sampled the image using regions of equal solid angle
using concentric rings about the
centre of the NS image.  This sampling method could be
performed due to the inherent symmetry in the circular NS
image.  By symmetry, the $s_2$ and $s_3$ components of the
polarization are zero.  Figure \ref{compare} shows the
frequency-dependent polarization signature for a NS with a
dipole moment of $\mu=10^{31}$G~cm$^3$.  The full-model polarization
(solid line) is plasma dominated at low frequencies and vacuum (QED)
dominated at higher frequencies.  In the transitory region, the total
polarization suffers some destructive interference between the two 
processes -- the total polarization is less than either the vacuum 
or plasma-dominated cases.

The location of the minimum in the transitory region can be understood
to be the frequency at which the photon decouples from the plasma and vacuum
effects at the same time.  Using the adiabatic condition,
\citet{Heyl:2000} determined the criteria for decoupling, 
\begin{equation}
\left|\hat{\Omega} \left( \frac{1}{|\hat{\Omega}|}\frac{\partial
|\hat{\Omega}|}{\partial r} \right)^{-1} \right| \approx 0.5
\end{equation}  

Using this expression, the distance at which decoupling occurs can be
solved for in both plasma-dominant and QED-dominant regimes.  To first
order these expressions are:
\begin{equation}
r_\rmscr{pl, qed} \propto \mu^{\frac{2}{5}} \omega^{\frac{1}{5}}
\times f(\alpha);\, 
r_\rmscr{pl, plas} \propto \left (\frac{\mu}{P \omega} \right)^{\frac{1}{2}} \times g(\alpha)
\end{equation}

Here $f(\alpha)$ and $g(\alpha)$ denote the functional dependence of
the expressions on the inclination of the star.  As the resonance
occurs when the photon decouples from both the vacuum and plasma
dominated effects at the same time, the critical frequency $\omega_c$
and the critical radius $r_c$ at which this resonance occurs are
easily calculated:
\begin{equation}
\omega_c \propto \frac{\mu^{\frac{1}{7}}}{P^{\frac{5}{7}}} \times h(\alpha);\,
r_c \propto \frac{\mu^{\frac{3}{7}}}{P^{\frac{1}{7}}} \times j(\alpha)
\end{equation}
The results of the numerical model were found to be in agreement with
these relations.

The critical polarization $P_c$ (the polarization at resonance) is
related to the to the radius at which photons decouple: 
$1- P_c \propto \left(R/r_c\right)^{\gamma}$
where $0 < \gamma \lesssim 2$. 

\begin{figure}
\psfrag{s1}{$s_1$}
\psfrag{ 1e+13}{$10^{13}$}
\psfrag{ 1e+14}{$10^{14}$}
\psfrag{ 1e+15}{$10^{15}$}
\psfrag{30m}{$30\mu$m}
\psfrag{3m}{$3\mu$m}
\psfrag{0.3m}{$0.3\mu$m}
\includegraphics[width=7.6cm,height=4.6cm]{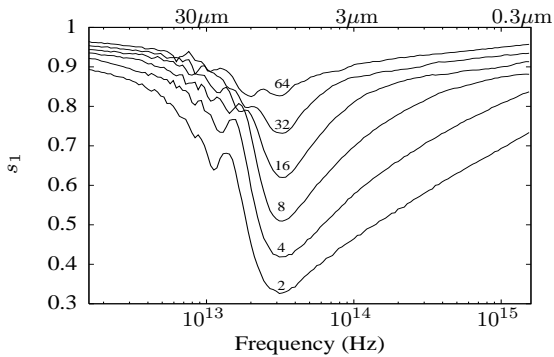}
\caption{\label{compB} Polarization signatures for various magnetic
field strengths, with period of 1.0s.  The plot shows polarization
signatures for increasing magnetic fields, with the lowest line
showing a field with a dipole moment of $2 \times 10^{31}$G~cm$^3$,
each subsequent line representing a factor of two increase in the
magnetic field to the last line, which shows the polarization
signature for a magnetic dipole moment of $6.4 \times
10^{32}$G~cm$^3$ 
Above each curve is the magnetic dipole moment in units of
$10^{31}$~G~cm$^3$.  In each case the dipole points
 30$^\circ$ from the line of sight. 
}
\end{figure}

\begin{figure}
\psfrag{s1}{$s_1$}
\psfrag{ 1e+13}{$10^{13}$}
\psfrag{ 1e+14}{$10^{14}$}
\psfrag{ 1e+15}{$10^{15}$}
\psfrag{30m}{$30\mu$m}
\psfrag{3m}{$3\mu$m}
\psfrag{0.3m}{$0.3\mu$m}
\includegraphics[width=7.6cm, height=4.6cm]{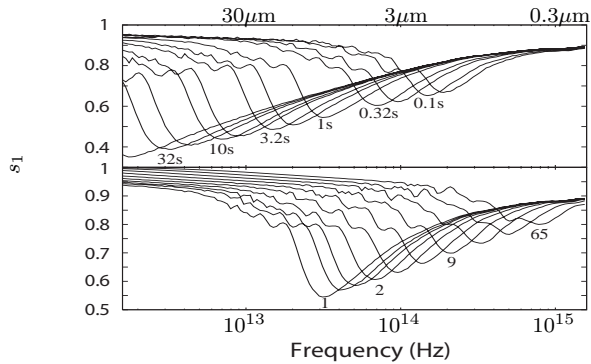}
\caption{\label{peri_plas}\label{period}\label{Plasma2}
Polarization signatures for a dipole moment of
$10^{31}$G cm$^3$ and various periods.  The upper panel shows the
polarization signatures for various NS periods.  The lines show,
starting from the most significant minimum on the left periods of 70s, 32s, 
20s,10s, 7s, 3.2s, 2s, 1s, 0.32s, 0.2s, 0.1s, and 0.07s.  Some of the
minima are labeled with the period. 
 The lower panel gives
the results for various total charge densities.
The lowest line shows the case in which the total charge
density is equal to the net charge density.  Each successive line above
represents an introduction of an additional component of the total charge
density proportional to the net charge density.  The second lowest
line represents an 25\% excess of charge, and each line
above that increases the excess by a factor of two. The minima are
labeled with the ratio of the total charge to the net charge.
 In each case the dipole points
 60$^\circ$ from the line of sight. }
\end{figure}

Figure \ref{compB} shows how the polarization signature changes as the
magnetic field strength is varied. It is observed that as the magnetic
field strength is increased, the strength of the resonance feature
decreases, but the critical frequency doesn't change significantly.
Varying the rotational period of the NS does change the
polarization signature.  The variation is due to the
effects of increased plasma density in the magnetosphere.  The results
seen in Figure \ref{period} show that as the rotational period is
changed the frequency of the dominant features are changed, but the
overall structure of the signature remains similar.

It is possible that the charge density of the plasma has been
underestimated.  \citet{Goldreich:1969} estimate only the net charge
density.  However, it is possible that the total charge density is
larger, as electron-positron pairs may exist in the magnetosphere.
Extending the equations of \citet{Mezaros:1992} into the formalism
presented here, the total charge density can be increased arbitrarily.
Figure \ref{Plasma2} shows that the introduction of additional charge
shifts the resonance to higher frequencies, and decreases the depth of
the feature.  Again, the overall shape of the feature remains similar.
Note that increasing the charge density is comparable to increasing
the rotational rate of the NS.  At these frequencies increasing the
total charge density is nearly covariant with increasing the spin rate
of the star, but the ellipticity of the modes depends only on the net
charge density so the results for higher net charge density
(increasing $\Omega$) and higher total charge density are slightly
different. 

\subsection{Phase Integrations}
\label{sec:phase-integrations}

\begin{figure}
\psfrag{blank}{Polarization}
\psfrag{stot}{$s_\rmscr{tot}$}
\psfrag{sO1}{$s_{O,1}$}
\psfrag{sO2}{$s_{O,2}$}
\includegraphics[width=7.6cm,height=4.6cm]{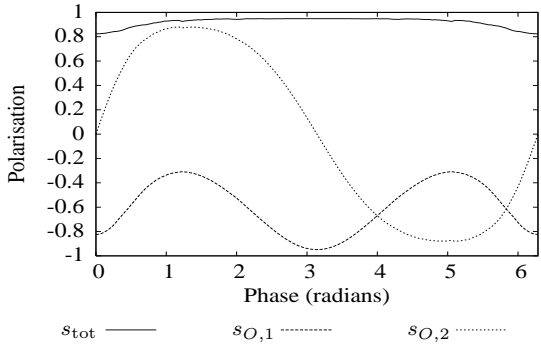}
\caption{\label{Single} Polarization from Neutron Star as a function
of phase for a NS with $\gamma = 30^{o}$ and $i_R = 30^\circ$,
for $\omega = 10^{13}$ rad/s and $\mu = 10^{31}$ G cm$^3$.  $s_\rmscr{tot}$
is the total polarization at the particular phase.  $s_{O,1}$ is the
component of polarization measured in direction perpendicular and
parallel to the rotational equator of the NS, and $s_{O,2}$ is the
component 45$^\circ$ to the equator.}
\end{figure}

It is currently impractical to measure the instantaneous polarization
of an object as faint as a neutron star; therefore it is important to
calculate the polarization signature from NS averaged over many spin
periods.  To do this the polarization must be calculated in a single
basis.  \citet{Heyl:2003} show how the
polarization can be calculated.  Define the latitude of the optical axis 
(observer) above rotational equator is as $i_R$ and the angle
between the magnetic axis and rotational axis as $\gamma$.
If the observer's polarimeter is aligned so that $s_{O,1}$ is
along the rotational equator of the NS and $s_{O,2}$ is defined
45$^\circ$ to it, the components of the polarization can be determined
from the calculations in the other bases\footnote[14]{The expression for
$s_{O,2}$ in \citet{Heyl:2003} is incorrect by a
factor of $\sin i_R$.  Note that the quadrature sum of $s_{O,1}$ and
$s_{O,2}$ should be unity.  This result is corrected here.  }
\begin{eqnarray}
s_{O,1} = s_1 \frac{2 \left( \sin \gamma \sin \phi \right) ^2}{1 -
\left( \cos\gamma \sin i_R + cos i_R \sin \gamma \cos \phi \right)^2}
-1\,  \\ 
s_{O,2} = s_1 \frac{2 \sin \gamma \sin \phi \left(\cos \gamma
\cos i_R - \cos \phi \sin \gamma \sin i_R \right)}{1 - \left( \cos
\gamma \sin i_R + \cos i_R \sin \gamma \cos \phi \right)^2}
\end{eqnarray}
The angle $\alpha$ as defined in the previous section can be
calculated using the law of cosines.
Figure~\ref{Single} shows how the polarization from a star evolves as
the phase varies.  The quadrature sum of the two terms adds to make
$s_\rmscr{tot}$, as expected.

\begin{figure}
\psfrag{sO1}{$s_{O,1}$}
\psfrag{ 1e+13}{$10^{13}$}
\psfrag{ 1e+14}{$10^{14}$}
\psfrag{ 1e+15}{$10^{15}$}
\psfrag{30m}{$30\mu$m}
\psfrag{3m}{$3\mu$m}
\psfrag{0.3m}{$0.3\mu$m}
\includegraphics[width=7.6cm,height=4.6cm]{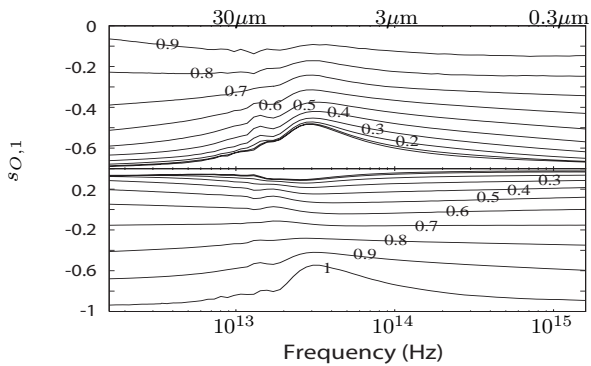}
\caption{\label{irg}
Phase-integrated polarization signatures.
The upper panel shows polarization signatures for various $i_R$ with
$\gamma = 30^\circ$.  
The lines show how the signature changes from
$\sin i_R = 0$ (bottom line) to $\sin i_R = 1.0$ (top line,
indistinguishable from the x-axis) with each above representing an
increase in $\sin i_R$ of 0.1.   The curves are labelled with the
value of $\sin i_R$.
The lower panel shows the results
for various $\gamma$ with
$i_R = 30^\circ$.  The lines show how the signature changes from $\cos
\gamma = 1.0 $ (bottom line) to $\cos \gamma = 0$ (top line) with each
above representing an decrement in $\cos \gamma$ of 0.1.  
The curves are labelled with the value of $\cos \gamma$.  In both
panels a
dipole moment of $10^{31}$G~cm$^3$ and an NS period of 1.0s were
assumed. }
\end{figure}

The phase-averaged polarization can be found by computing the mean values
in $s_{O,1}$ and $s_{O,2}$ with the average value of $s_{O,2}$ being zero by
symmetry.  The lower panel of Figure~\ref{irg} shows how the signature changes as $i_R$
varies.  Observe that when $\sin i_R = 1$ (top line) -- the observer is
looking directly down on the rotation axis -- there is no observed
polarization.  However, for observations at lower latitudes there is a
significant net polarization and a resonant feature at approximately
$3 \times 10^{13}$ Hz.  Statistically, half of all NS will
have inclination angles such that $ \sin i_R \leq 0.5$.  All the
models presented in this range show strong features in this range
considered.


The upper panel of Figure~\ref{irg} shows how the signature changes as
$\gamma$ varies.  The strongest resonant feature occurs when the
magnetic and rotational axes are aligned ($\gamma=0$).  However,
strong resonant features can be observed for $\cos \gamma \ge 0.5$.


\section{Discussion}

Of particular attention is the wavelength at which the resonant
feature occurs - the near infrared.  As this energy, it would be
possible to measure the polarization of the thermal radiation from a
NS using a large ground-based telescope or Hubble's NICMOS instrument.
This instrument is only sensitive to wavelengths less than 2.1~$\mu$m,
so it is possible that a large portion of the behaviour could not be
probed.  If the total plasma density is significantly larger than the
Goldreich-Julian plasma density, the feature moves into near infrared
or optical energies, and more observing options become available.

This letter presents a model for the evolution of polarized light
through the magnetosphere of highly magnetized neutron stars. We
found that the polarization has a frequency-dependent
signature, where the total polarization decreases significantly in the
optical or near IR, dependent on the plasma density near the neutron
star and the structure of the magnetic field. Observations of neutron
stars in this frequency region may provide useful probes of
the properties of neutron-star magnetospheres and elucidate the processes
responsible for pulsar emission.

\section*{Acknowledgments}

The Natural Sciences and Engineering Research Council of Canada
supported this work.  Correspondence and requests for materials should
be addressed to J.S.H.(heyl@phas.ubc.ca).

\bibliographystyle{mn2e}
\bibliography{thesis}

\label{lastpage}

\end{document}